\documentclass[12pt]{article} 
\usepackage{amssymb,epsfig,ldd_art}
\newcommand{\One}{1\kern-4.5pt1}
\newcommand{\lapprox}{\raisebox{-0.5ex}{$\ 
\stackrel{\textstyle<}{\textstyle\sim}\ $}}
\begin{document}

\addtolength{\baselineskip}{0.20\baselineskip}

\rightline{SWAT/03/382}

\hfill October 2003


\vspace{48pt}

\centerline{\Large High Density Effective Theory Confronts the Fermi Liquid}


\vspace{18pt}

\centerline{\bf Simon Hands}

\vspace{15pt}

\centerline{{\sl Department of Physics, University of Wales Swansea,}}
\centerline{\sl Singleton Park, Swansea SA2 8PP, U.K.}
\smallskip

\vspace{24pt}


\centerline{{\bf Abstract}}

\noindent
{\narrower 
The high density effective theory recently introduced by Hong and Hsu
to describe
ultradense relativistic fermionic matter is used
to calculate the tree-level
forward scattering amplitude between two particles at the Fermi
surface. While the direct term correctly reproduces that of the underlying gauge
theory, the exchange term has the wrong sign. The physical
consequences are discussed in the context of Landau's
theoretical description  of the Fermi liquid.
}




\vfill
\newpage
\section{Introduction}

When studying any strongly-interacting system, an effective description which
highlights the low energy degrees of freedom is highly desirable. Such a
programme has recently been initiated by Hong and collaborators for the case
of systems with a large number density of quarks, 
in which case the simplified description keeps as
excitations about the ground state just those particle and hole states in the 
immediate vicinity of the Fermi surface. The resulting high density effective 
theory (HDET) \cite{Hong2} 
is derived from QCD in very much the same way as the heavy
quark effective theory (HQET) \cite{HQET}, 
which has proved so successful in the description of
bound states of heavy and light quarks. In HQET, heavy quarks of mass $M$
moving with
4-velocity $v_\mu$ are described by fields $h_v$ defined in terms of the 
original quark fields by 
\begin{equation}
h_v(x)=e^{iMv^\mu x_\mu}P_v\psi(x)\;\;\;\mbox{with}\;\;\;
P_v={\textstyle{1\over2}}(1+v{\!\!\!/\,}).
\end{equation}
The projector $P_v$ enables the kinetic term for the $h_v$ fields to be 
independent of the large scale $M$, while
the phase factor ensures that even when the $h_v$ fields are restricted to
carry some residual momentum $p\ll M$, 
gluons which scatter off the heavy quark
exchange momentum with a state carrying physical momentum $P\simeq O(M)$.
HDET employs a similar manipulation with $M$ replaced by the chemical
potential $\mu$, which for a degenerate system at zero temperature may be
identified with the Fermi energy. It is then possible to devise a theory
in which both particle and hole excitations
appear with equal measure in the path integral \cite{HH1}.

The HDET approach has succeeded in calculating the screening mass of electric
gluons in quark matter, and in calculating the strength of the gap induced by 
a BCS instability to diquark pairing in the color-flavor locked (CFL) channel
\cite{Hong2}. Most intriguingly, in Euclidean metric it yields at leading order
in $\Lambda/\mu$ (where $\Lambda$ is the low energy scale of interest) a
positive definite path
integral measure \cite{HH1}. This has enabled a Vafa-Witten style proof that the
CFL phase is the true ground state of QCD at asymptotically high density
\cite{HH2}, as well as raising the possibility that HDET might form a starting
point for non-perturbative lattice simulations of dense matter.

There is, however, one important physical respect in which dense matter differs
from the heavy-light systems described by HQET. In the latter case the only
important interactions of the heavy quarks are 
small-angle scattering by $t$-channel exchange of gluons with light quarks.
In quark matter scattering takes place between identical particles, implying
that interactions in the $u$-channel, in which large momenta
may be exchanged between two non-parallel particles at the Fermi surface,
 are also relevant. In many-body and nuclear
physics these two distinct contributions are known respectively
as ``direct'' and
``exchange'' processes. While HQET and HDET both capture the essential physics
of the direct interaction, for HDET one should also check its behaviour
in the exchange channel.

After reviewing the version of HDET as set out in Refs.~\cite{Hong2,HH1} in the next section, 
I will outline in Sec.~\ref{sec:forward} an explicit calculation of the
amplitude for forward scattering between two particle states at the Fermi
surface, using both gauge theory and HDET, and show that while HDET successfully
reproduces the direct term, there is a mismatch for the exchange. Forward
scattering is interesting for degenerate systems, and indeed, is known to be the
only non-irrelevant 
interaction at the Fermi surface in the renormalisation group sense
\cite{relevance}. In
Sec.~\ref{sec:fermi} I will show that this amplitude is a central feature
in an alternative and much older phenomenological description of dense matter,
the Fermi liquid \cite{Landau}. 
The consequences of an incorrect description of
exchange interactions for the detailed 
relations between Fermi liquid parameters such as the Fermi
energy, momentum and velocity, as well as for the behaviour of collective
excitations, are then set out in
Sec.~\ref{sec:confront}. Finally I present some brief conclusions.

\section{High Density Effective Theory}
We begin by reviewing the derivation of the high density effective theory
(HDET), starting from the fermionic part of the gauge theory Lagrangian density
and including a chemical potential $\mu$
(we neglect any bare quark mass):
\begin{equation}
{\cal L}=\bar\psi(x)(i\gamma^\mu D_\mu+\mu\gamma_0)\psi(x)
\label{eq:start}
\end{equation}
where the covariant derivative $D_\mu\equiv\partial_\mu+igA_\mu$.
Rewrite the physical 
quark 3-momentum as ${\bf k}=\mu\hat{\bf p}+{\bf p}$ where the
``residual'' momentum, giving the distance from the Fermi surface, satisfies ${\bf
p}\parallel{\bf k}$. HDET assumes $\vert{\bf p}\vert\ll\mu$, and is expressed in
terms of fields most naturally thought of in reciprocal
space as functions of {\bf p}. This is achieved via the decomposition into
``fast'' $\psi_-$ and ``slow'' $\psi_+$ degrees of freedom
\begin{equation}
\psi(x)=\exp(i\mu{\bf x}.\hat{\bf p})[\psi_+(x)+\psi_-(x)]
\label{eq:decomp}
\end{equation}
where $\vert\hat{\bf p}\vert=1$ and
$\psi_\pm$ are eigenstates of the projection operators
\begin{equation}
P_\pm(p)={\textstyle{1\over2}}(1\pm\vec\alpha.\hat{\bf p}).
\end{equation} 
The matrix $\vec\alpha=\gamma_0\vec\gamma$, and where needed we use the
representation
\begin{equation}
\vec\alpha=\left(\matrix{&\vec\sigma\cr\vec\sigma&\cr}\right).
\end{equation}
We also define a new covariant derivative $\tilde D_\mu$:
\begin{equation}
\tilde D_\mu
=\partial_\mu+ige^{-i\mu{\bf x}.\hat{\bf p}}A_\mu e^{i\mu{\bf x}.\hat{\bf
p}}\equiv\partial_\mu+ig\tilde A_{\mu}.
\label{eq:Adecomp}
\end{equation}
In real space the fields must be constructed using the operator
$\hat{\bf p}=(-i/\surd\nabla^2)\vec\nabla$.
In terms of the new fields the quark Lagrangian (\ref{eq:start}) becomes
\begin{eqnarray}
{\cal L}&=&\bar\psi_+\gamma_0(1,\hat{\bf
p})^\mu i\tilde D_\mu\psi_+
+\bar\psi_-\gamma_0[(1,-\hat{\bf
p})^\mu i\tilde D_\mu
+2\mu]\psi_-
+\left[\bar\psi_-i\gamma^\mu_\perp\tilde D_\mu\psi_++h.c.\right]\nonumber\\
&=&\bar\psi_+i\gamma^\mu_\parallel\tilde D_\mu\psi_+
+\bar\psi_-(i\gamma^\mu_\parallel\tilde D_\mu+
2\mu\gamma_0)\psi_-
-g\left[\bar\psi_-\gamma^\mu_\perp\tilde A_\mu\psi_++h.c.\right]
\label{eq:rewrite}
\end{eqnarray}
where we define $\gamma^\mu_\parallel=(\gamma_0,\hat{\bf p}\vec\gamma.\hat{\bf
p})^\mu$, $\gamma^\mu_\perp=\gamma^\mu-\gamma^\mu_\parallel$, and the form of
the final term follows from
$\gamma^\mu_\parallel\partial_\mu\equiv\partial{\!\!\!/\,}$,
$\gamma^\mu_\perp\partial_\mu=0$.
We have used the identities \cite{Hong2}
\begin{equation}
P_\mp\gamma^\mu P_\pm=\gamma_0(1,\pm\hat{\bf p})^\mu P_\pm\;\;\;;\;\;\;
P_\pm\gamma^\mu P_\pm=\gamma^\mu_\perp P_\pm.
\end{equation}
For $p\ll\mu$ it is possible to integrate out the fast modes $\psi_-$ to yield a
low energy effective Lagrangian written solely in terms of $\psi_+$. At tree
level this is done via the equation of motion
\begin{equation}
\psi_-={{g\gamma_0}\over{2\mu+i\overline{D}_\parallel}}\gamma^\mu_\perp
\tilde A_\mu\psi_+={{g\gamma_0}\over{2\mu}}\sum_{n=0}^\infty
\left(-{{i\overline{D}_\parallel}\over{2\mu}}\right)^n\gamma^\mu_\perp
\tilde A_\mu\psi_+
\end{equation}
where $\overline{D}_\parallel=(1,-\hat{\bf p})^\mu\tilde D_\mu$. We obtain
the HDET effective Lagrangian as a derivative expansion, ie. in powers of
$\Lambda/\mu$ where $\Lambda\ll\mu$ denotes a momentum scale of physical interest:
\begin{equation}
{\cal L}_{HDET}=\bar\psi_+i\gamma^\mu_\parallel\tilde D_\mu\psi_+
+{g^2\over{2\mu}}\bar\psi_+\gamma_0(\gamma^\mu_\perp
\tilde A_\mu)^2\psi_++O\left({D^2\over\mu^2}\right).
\label{eq:Lhdet}
\end{equation}
The leading order term of 
HDET should therefore be able to describe the dynamics of
low energy excitations near the Fermi surface, in which any momentum
transferred by scattering off a gluon should be much smaller than the Fermi
momentum. At higher order, further terms in the HDET action such as a
four-fermi contact 
interaction and a gluon screening mass are generated
\cite{Hong2}. 

In fact, HDET has to be subtly modified in order to preserve unitarity. The
$\psi_-$ excitations are nothing other than anti-particles, whose physical
impact in superdense matter with $p\ll\mu$ is relatively unimportant. However, 
just as in a metal or a semiconductor, we must also take hole excitations in the
Fermi sea into account, which have identical quantum numbers to anti-particles,
but play a much more important role.
In order to do this consistently we modify the definitions (\ref{eq:decomp},
\ref{eq:Adecomp})
to read
\cite{HH1}:
\begin{equation}
\psi(x)=e^{iX}\psi_+(x)\;\;\;;\;\;\;\tilde A_\mu(x)=e^{-iX}A_\mu(x)e^{iX}
\;\;\;\mbox{with}\;\;\;
X\equiv\mu({\bf x}.\hat{\bf p})(\vec\alpha.\hat{\bf p}).
\label{eq:unit}
\end{equation}
Note that $\psi_+(p)$ can now satisfy either of $P_\pm(p)\psi_+=\psi_+$ depending
on whether the physical state it describes is a particle or a hole, in either
case with momentum {\bf p}.
It is not possible, however, to derive HDET from the original gauge theory
starting with the projection (\ref{eq:unit}).
The quark-gluon vertex
arising from the leading order term
$\bar\psi_+i\gamma^\mu_\parallel\tilde D_\mu\psi_+$ in ${\cal L}_{HDET}$
now reads 
\begin{equation}
{\cal L}_{qqg}=-g\int_x A^\mu(q)e^{iqx}
\bar\psi_+(p^\prime)e^{-ip^\prime x}
e^{-i\mu{\bf x}.\hat{\bf p}^\prime\vec\alpha.\hat{\bf p}^\prime}
(\gamma_0,\hat{\bf p}.\vec\gamma\hat{\bf p})_\mu
e^{ipx}e^{i\mu{\bf x}.\hat{\bf
p}\vec\alpha.\hat{\bf p}}\psi_+(p).
\end{equation}
After integrating over $x$ we find
\begin{eqnarray}
{\cal L}_{qqg}&=&-gA^\mu(q)\Bigl[
\bar\psi_+(p^\prime)P_-(p^\prime)\gamma_0(1,+\hat{\bf p})_\mu
P_+(p)\psi_+(p)
\delta^4\Bigl((p^\prime+\mu\hat{\bf p}^\prime)-(p+\mu\hat{\bf p})-q\Bigr)
\nonumber\\
&+&\bar\psi_+(p^\prime)P_+(p^\prime)
\gamma_0(1,-\hat{\bf p})_\mu
P_-(p)\psi_+(p)
\delta^4\Bigl((p^\prime-\mu\hat{\bf p}^\prime)-(p-\mu\hat{\bf p})-q\Bigr)
\nonumber\\
&+&\bar\psi_+(p^\prime)P_-(p^\prime)
\gamma_0(1,-\hat{\bf p})_\mu
P_-(p)\psi_+(p)
\delta^4\Bigl((p^\prime+\mu\hat{\bf p}^\prime)-(p-\mu\hat{\bf p})-q\Bigr)
\nonumber\\
&+&\bar\psi_+(p^\prime)P_+(p^\prime)
\gamma_0(1,+\hat{\bf p})_\mu
P_+(p)\psi_+(p)
\delta^4\Bigl((p^\prime-\mu\hat{\bf p}^\prime)-(p+\mu\hat{\bf p})-q\Bigr)
\Bigr],
\label{eq:vertex}
\end{eqnarray}
the four terms describing respectively particle or a hole scattering off a
gluon, and
particle-hole creation and annihilation.
The effect of the phase factor in (\ref{eq:unit}) has been to ensure that the
gluon scatters off states carrying physical rather than residual momenta.
The Dirac structure in principle is given by expressions such as
\begin{equation}
P_-(p^\prime)\gamma_0P_+(p)=
{\gamma_0\over4}\left[
1+\hat{\bf p}^\prime.\hat{\bf p}+\vec\alpha.(\hat{\bf p}^\prime+\hat{\bf p})
+i\gamma_5\vec\alpha.\hat{\bf p}^\prime\times\hat{\bf p}\right]
\label{eq:verfac}
\end{equation}
Note, however, that since by assumption $\hat{\bf p}^\prime.\hat{\bf
p}=1+O(p^2/\mu^2)$, $\hat{\bf p}^\prime\times\hat{\bf p}=O(p/\mu)$, at leading
order in HDET we can
consistently approximate
\begin{equation}
P_\mp(p^\prime)\gamma_0P_\pm(p)\approx\gamma_0
P_\pm(p).
\end{equation}
In this same limit $P_\pm(p^\prime)\gamma_0P_\pm(p)\approx0$.

In Euclidean metric, the leading term of the HDET Lagrangian (\ref{eq:Lhdet})
is of the form $\bar\psi_+M\psi_+$ where $M$ is both anti-hermitian and
anticommutes with $\gamma_5$, implying that the functional measure of the path
integral $\mbox{det}M$ is real and positive. This raises the attractive
possibility of using the HDET action as a basis for non-perturbative simulations
of systems at asymptotically high quark densities
using lattice Monte Carlo methods \cite{HH1,HH2}. 
The higher order terms in (\ref{eq:Lhdet}) describing corrections of
$O(\Lambda/\mu)$
are of the form $\bar\psi_+H\psi_+$ where $H$ is hermitian. In general,
$\mbox{det}(M+H)$ is complex, leading to a restoration of the notorious ``sign
problem'' at finite density. It may prove possible, however, to find means to
treat
corrections for $\Lambda/\mu\lapprox 1$, much as corrections to QCD simulations at
$\mu=0$ can be calculated for $\mu/T\lapprox1$ \cite{muonT}.

\section{Forward Scattering}
\label{sec:forward}

\begin{figure}[htb]
\bigskip\bigskip
\begin{center}
\epsfig{file=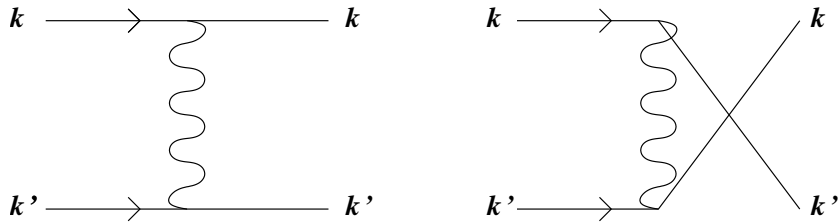, width=11cm}
\end{center}
\caption{Feynman diagrams for forward scattering, showing direct (left) and
exchange (right) contributions.}
\label{fig:Feynman}
\end{figure}

Let us consider a very simple process, tree-level 
forward scattering of two
relativistic particles at the Fermi surface, by
comparing calculations made using both HDET and the full underlying quantum
field theory, which for simplicity we will initially assume to be QED.
The relevant Feynman diagrams are shown in Fig.~\ref{fig:Feynman}.
In Minkowski metric and Feynman gauge
the QED scattering amplitude for the direct process is 
\begin{equation}
{\cal A}^{dir}(k,k^\prime)
={1\over{4\varepsilon_{\bf k}\varepsilon_{{\bf k}^\prime}}}{\cal M}^{dir}({\bf
k},{\bf k}^\prime)
\end{equation}
with 
\begin{equation}
i{\cal M}^{dir}=ie^2{g^{\mu\nu}\over{-\lambda^2}}
\bar u(k,s)\gamma_\mu u(k,s)\bar u(k^\prime,s^\prime)\gamma_\nu
u(k^\prime,s^\prime).
\end{equation}
Here, $e$ is the electron charge, 
an infra-red regulator mass $\lambda$ has
been introduced for the photon, and $u$, $\bar u$ are plane-wave spinors
normalised to $\bar uu=2m$, with $m$ the electron mass. 
The four-momentum $k_\mu=(\varepsilon_{\bf k},
{\bf k})_\mu$, with 
the energy
$\varepsilon_{\bf k}$ of course equal to 
$\mu$ at the Fermi surface. 

To evaluate the
spin-symmetric part of the amplitude we average over $s,s^\prime$ to obtain
\begin{equation}
{\cal A}^{dir}(k,k^\prime)=-{1\over{16\varepsilon_{\bf k}\varepsilon_{{\bf
k}^\prime}}}{e^2\over{\lambda^2}}
\mbox{tr}[\gamma_\mu(k{\!\!\!/\,}+m)]\mbox{tr}[\gamma^\mu(k{\!\!\!/\,}^\prime+m)]
=-{e^2\over\lambda^2}{{\varepsilon_{\bf k}\varepsilon_{{\bf k}^\prime}-
{\bf k}.{\bf
k}^\prime}\over{\varepsilon_{\bf k}\varepsilon_{{\bf k}^\prime}}}.
\end{equation}
The exchange contribution includes a relative minus sign due to Fermi
statistics:
\begin{eqnarray}
{\cal A}^{ex}(k,k^\prime)&=&-{1\over{16\varepsilon_{\bf k}\varepsilon_{{\bf
k}^\prime}}}{e^2\over{(k-k^\prime)^2-\lambda^2}}\mbox{tr}
[\gamma_\mu(k{\!\!\!/\,}^\prime+m)\gamma^\mu(k{\!\!\!/\,}+m)]\nonumber\\
&=&{1\over{2\varepsilon_{\bf k}\varepsilon_{{\bf
k}^\prime}}}{e^2\over{({\bf k}-{\bf k}^\prime)^2+\lambda^2}}
[-(\varepsilon_{\bf k}\varepsilon_{{\bf k}^\prime}
-{\bf k}.{\bf k}^\prime)+2m^2].
\end{eqnarray}
Note that $k_0=k_0^\prime$ for two states at the Fermi surface.
In the limit $m,\lambda\to0$ 
we have $\varepsilon_{\bf k}=\vert{\bf k}\vert=\mu$ and \cite{BC}
\begin{eqnarray}
{\cal A}_{QED}^{dir}(k,k^\prime)&=&-{e^2\over\lambda^2}(1-\cos\theta),
\label{eq:dirqed}\\
{\cal A}_{QED}^{ex}(k,k^\prime)&=&-{e^2\over4\mu^2},
\label{eq:qed}
\end{eqnarray}
where $\theta$ is the angle between the two particle momenta.

To repeat the calculation using HDET 
a necessary ingredient are the plane wave states.
Eigenstates $\Psi_\pm$ of the momentum space HDET Dirac equation 
\begin{equation}
p_0\Psi_\pm(p)=\vec\alpha.{\bf p} P_\pm(p)\Psi_\pm(p)
\end{equation}
are given by
\begin{eqnarray}
\Psi_+(p,s)=\left(\matrix{\phantom{+\vec\sigma.\hat{\bf p}}\phi^s\cr
+\vec\sigma.\hat{\bf p}\,\phi^s\cr}\right)\;\mbox{with
eigenvalue}\;p_0=+\vert{\bf p}\vert;\nonumber \\
\Psi_-(p,s)=\left(\matrix{\phantom{+\vec\sigma.\hat{\bf p}}\chi^s\cr
-\vec\sigma.\hat{\bf p}\,\chi^s\cr}\right)\;\mbox{with
eigenvalue}\;p_0=-\vert{\bf p}\vert.
\end{eqnarray}
The two-spinors $\phi,\chi$ are given by
$\phi^1=\chi^2={\scriptstyle\left(\matrix{1\cr0\cr}\right)}$, 
$\phi^2=\chi^1={\scriptstyle\left(\matrix{0\cr1\cr}\right)}$, whence
\begin{equation}
\Psi_\pm^\dagger\Psi_\pm=2\;\;\;;\;\;\;\bar\Psi_\pm\Psi_\pm=0\;\;\;;\;\;\;
\sum_s\Psi_\pm(p,s)\bar\Psi_\pm(p,s)=\gamma_\mu(1,\pm\hat{\bf p})^\mu.
\label{eq:norm}
\end{equation}
Note also the important identity
\begin{equation} \Psi_\pm(p,s)\equiv\Psi_\mp(-p,-s).\label{eq:id}
\end{equation}

Now let's use HDET, 
and in particular the $qqA$ vertex described by the first term
of (\ref{eq:vertex}),  to calculate the spin-symmetric 
forward scattering amplitude between two
particles at the Fermi surface. First we examine the direct term; with the 
normalisation (\ref{eq:norm}) we have
\begin{eqnarray}
{\cal A}_{HDET}^{dir}(p,p^\prime)&=&{1\over16}\sum_{s,s^\prime}
{\cal M}^{dir}(p,p^\prime)\nonumber\\
&=& {e^2\over16}{g^{\mu\nu}\over{-\lambda^2}}
\sum_{s,s^\prime}\bar\Psi_+(p,s)\gamma_0(1,+\hat{\bf p})_\mu\Psi_+(p,s)
\bar\Psi_+(p^\prime,s^\prime)\gamma_0(1,+\hat{\bf
p}^\prime)_\nu\Psi_+(p^\prime,s^\prime)\nonumber\\
&=&
-{e^2\over{16\lambda^2}}\mbox{tr}[\gamma_0\gamma_\rho]
\mbox{tr}[\gamma_0\gamma_\sigma]
(1,\hat{\bf p})^\rho(1,\hat{\bf p}^\prime)^\sigma(1-\hat{\bf p}.\hat{\bf
p}^\prime)={\cal A}_{QED}^{dir}(p,p^\prime).
\end{eqnarray}
The direct term calculated in HDET thus agrees with that of the full theory
(\ref{eq:dirqed}).
For the exchange term,
\begin{equation}
{\cal
A}_{HDET}^{ex}(p,p^\prime)=-{1\over16}\sum_{s,s^\prime}{e^2\over{q^2-\lambda^2}}
\bar\Psi_+(p^\prime,s^\prime)\gamma_0(1,\hat{\bf
p})^\mu\Psi_+(p,s)\bar\Psi_+(p,s)\gamma_0(1,\hat{\bf
p}^\prime)_\mu\Psi_+(p^\prime,s^\prime),
\end{equation}
with ${\bf q}=(\vert{\bf p}^\prime\vert+\mu)\hat{\bf p}^\prime-(\vert{\bf
p}\vert+\mu)\hat{\bf p}$ (as before, for two states near
the Fermi surface $q_0\approx0$). 
Performing the average over spins and resulting trace
we find
\begin{equation}
{\cal A}_{HDET}^{ex}(p,p^\prime)
=-{e^2\over{4q^2}}[2g_{\rho 0}g_{\sigma 0}-g_{00}g_{\rho\sigma}](1,\hat{\bf
p})^\rho(1,\hat{\bf p}^\prime)^\sigma(1-\hat{\bf p}.\hat{\bf p}^\prime)
=-{e^2\over{4q^2}}(1-\cos^2\theta).
\end{equation}
Finally, $q^2=-2\mu^2(1-\cos\theta)(1+O(p/\mu))$, so that to leading order the 
result is
\begin{equation}
{\cal A}_{HDET}^{ex}(p,p^\prime)={e^2\over{8\mu^2}}(1+\cos\theta).
\label{eq:ahdet}
\end{equation}
If the calculation is repeated with the more accurate 
vertex factor (\ref{eq:verfac}), then the answer is
\begin{equation}
{\cal A}_{HDET}^{ex}(p,p^\prime)={e^2\over{8\mu^2}}
\left[{{(1+\cos\theta)^2}\over2}+{1\over8}\sin^2\theta(1-\cos\theta)\right].
\end{equation}
As anticipated, this result agrees with (\ref{eq:ahdet}) up to terms of
$O(p^2/\mu^2)$.
In either case, though,
 the HDET result, though of the same order of magnitude, actually
disagrees even in overall sign with that of QED
(\ref{eq:qed}). 

Since QED and HDET disagree even at leading order in $p/\mu$, it is
difficult to see how the discrepancy between (\ref{eq:qed}) and
(\ref{eq:ahdet}) can be sorted out at higher order; indeed, the next term 
$(2\mu)^{-1}\bar\psi_+\gamma_0(e\tilde A{\!\!\!/\,}_\perp)^2\psi_+$ 
in the derivative
expansion would correct ${\cal A}^{ex}_{HDET}$ only at $O(e^4)$. In fact, to
reconcile the two approaches we need to step back to the mixed term in
(\ref{eq:rewrite})
before $\psi_-$ is eliminated: 
$e[\bar\psi_-\tilde A{\!\!\!/\,}_\perp\psi_++h.c.]$.
This yields a vertex
$-ie\gamma_0(0,\vec\alpha-\hat{\bf p})_\mu$, which can contribute to particle
-- particle scattering once we realise that an anti-particle created by
$\bar\psi_-(p)$ is indistinguishable from a particle with momentum $-p$.
The new vertex does not contribute to direct scattering, but does yield
three
additional contributions to ${\cal A}^{ex}$ at $O(e^2)$, eg:
\begin{eqnarray}
{\cal A}^{ex}_{+-}&=&-{e^2\over{16q^2}}\sum_{s,s^\prime}
\bar\Psi_+(p^\prime,s^\prime)\gamma_0(1,\hat{\bf p})^\mu\Psi_+(p,s)
\bar\Psi_-(-p,-s)\gamma_0(0,\vec\alpha-\hat{\bf
p}^\prime)_\mu\Psi_+(p^\prime,s^\prime)\nonumber\\
&=&-{e^2\over{4q^2}}\biggl[-p_i^\prime(g_{\rho 0}g_{\sigma i}+g_{\rho
i}g_{\sigma 0})+\hat{\bf p}.\hat{\bf p}^\prime(2g_{\rho 0}g_{\sigma
0}-g_{\rho\sigma}g_{00})\biggr](1,\hat{\bf p}^\prime)^\rho(1,\hat{\bf
p})^\sigma\nonumber\\
&=&{e^2\over{8\mu^2}}{{(1+\cos\theta)^2}\over{1-\cos\theta}},
\end{eqnarray}
where use has been made of (\ref{eq:id}). We then find that to $O(e^2)$
\begin{equation}
{\cal A}^{ex}_{++}+
{\cal A}^{ex}_{+-}+
{\cal A}^{ex}_{-+}+
{\cal A}^{ex}_{--}=
{\cal A}^{ex}_{QED}
\end{equation}
where $++$ denotes the original HDET term (\ref{eq:ahdet}). We conclude that
once the $\psi_-$ degrees of freedom have been  eliminated to define the effective
theory solely in terms of $\psi_+$, the exchange amplitude (\ref{eq:qed}) cannot
be faithfully reproduced at tree level.

\section{The Fermi Liquid}
\label{sec:fermi}

Why is this important? The forward scattering amplitude plays a central role in
another, much older phenomenological description of degenerate matter, 
the Fermi liquid \cite{Landau, Landau2}. The essential physical idea is that the
dominant low-energy excitations in the neighbourhood of the Fermi surface are
quasiparticle states having energy $\varepsilon_{\bf k}$, width of
$O(\varepsilon_{\bf k}-\mu)^2$ and equilibrium distribution
\begin{equation}
n_{\bf k}={1\over{\exp\left({{\varepsilon_{\bf k}-\mu}\over T}\right)+1}}.
\end{equation}
For temperature $T=0$ we expect $\varepsilon_{\bf k}$ to have the form
\begin{equation}
\varepsilon_{\bf k}\simeq\mu+\beta_F(\vert{\bf k}\vert-k_F)
\end{equation}
where $k_F$, $\beta_F\equiv\vert\vec\nabla_{\bf k}\varepsilon_{\vert{\bf
k}\vert=k_F}\vert$ 
are respectively the Fermi momentum and Fermi velocity, which at
this level are
phenomenological parameters. Leading order HDET assigns them 
the free-field values $k_F=\mu$, $\beta_F=1$. Fermi liquid
 theory derives quantitative
power via the equation for the variation in quasiparticle energy in response to
a departure $\delta n$ from equilibrium:
\begin{equation}
\delta\varepsilon_{\bf k}=\int{{d^3k^\prime}\over{(2\pi)^3}}
{\cal F}_{{\bf k},{\bf k}^\prime}\delta n_{{\bf k}^\prime}.
\end{equation}
The Fermi liquid interaction ${\cal F}_{{\bf k},{\bf k}^\prime}$ is given to
lowest order in perturbation theory by \cite{Landau3,AGD}
\begin{equation}
{\cal F}_{{\bf k},{\bf k}^\prime}=-{\cal A}({\bf k},{\bf k}^\prime)=
-{\cal A}(\hat{\bf p},\hat{\bf p}^\prime)
\label{eq:fl}
\end{equation}
where in the last step we have assumed that the states are so close to the Fermi
surface that the matrix element only depends on the relative orientation of the
particle momenta. If ${\cal F}$ can be calculated in some scheme, it is then
possible to obtain quantitative relations between the parameters $\mu$, $k_F$
and $\beta_F$, as well as derive the velocities of collective excitations
\cite{Landau, BC, Landau2}. 

\begin{figure}[htb]
\bigskip\bigskip
\begin{center}
\epsfig{file=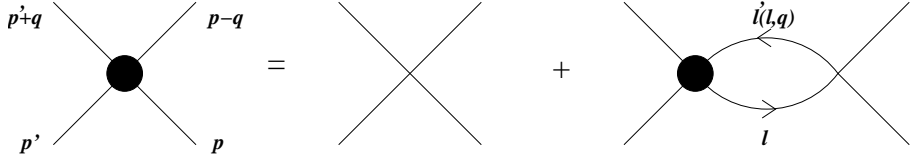, width=12cm}
\end{center}
\caption{Schwinger-Dyson equation for the scattering amplitude
$-\Gamma(p^\prime,p,q)$.}
\label{fig:sdzero}
\end{figure}
It may be helpful to outline the origin of the relation (\ref{eq:fl}) -- to my
knowledge there
is no argument more elegant than the original one presented by Landau in one of
the earliest papers on non-perturbative field theory \cite{Landau3,AGD}.
Consider quasiparticle scattering with the kinematics sketched in
Fig.~\ref{fig:sdzero}: the full amplitude is then given by
\begin{equation}
\Gamma(p^\prime,p,q)=\Gamma^0(p^\prime,p)-i\int{{d^4\ell}\over{(2\pi)^4}}
\Gamma(p^\prime,\ell,q)S(\ell)S(\ell^\prime(\ell,q))\Gamma^0(q,p).
\label{eq:sdzero}
\end{equation}
We wish to focus on the singular behaviour of $\Gamma$ as the momentum
transfer $q\to0$, so separate
out the part $\Gamma^0$ (shown as a simple point vertex in
Fig.~\ref{fig:sdzero}) which is regular in this limit. The singular behaviour
results from the quasiparticle propagator product $S(\ell)S(\ell^\prime)$, and
in particular from the contribution of a particle-hole
pair, in which no net quark number flows through the
diagram, for loop momenta on the Fermi
surface. Using HDET propagators which are functions of residual rather than
physical momenta, but with spinor indices suppressed
for simplicity, this product contains terms like
\begin{equation}
S(\ell)S(\ell^\prime(\ell,q))={\mathfrak g}\times
{1\over{\ell_0-\vert{\bf l}\vert-i\epsilon}}\;\; 
{1\over{(\omega-\ell_0)
+(\vert{\bf l}\vert-\vert{\bf q}\vert\cos\theta)+i\epsilon}} 
\end{equation}
where we have written $\omega=-q_0$, the $i\epsilon$ terms are chosen so that
the first factor descibes a particle of momentum $\ell$, the second a hole of 
momentum $-\ell^\prime$, and the kinematics enforced by eg. 
the $\delta$-functions
of the third or fourth terms of (\ref{eq:vertex}) are such that 
$\vert{\bf l}^\prime\vert=-\vert{\bf l}\vert+\vert{\bf q}\vert\cos\theta$, 
where $\theta$ is the angle between {\bf q} and {\bf l} and we
have assumed all momenta $\ll\mu$ so that the curvature of the Fermi surface
can be neglected and ${\bf l}^\prime\parallel{\bf l}$.
The factor ${\mathfrak g}$ arises from a trace over internal degrees of
freedom, and counts the degeneracy of each momentum state.
Within the integral in (\ref{eq:sdzero}) the product may be approximated by 
\begin{equation}
S(\ell)S(\ell^\prime)\simeq 2\pi{\mathfrak g}i
{{\vert{\bf q}\vert\cos\theta}\over
{\omega-\vert{\bf q}\vert\cos\theta}}\delta(\ell_0)\delta(\vert{\bf l}\vert)
+G(\ell)
\end{equation}
where the pole structure has resulted in the first term being supported only 
for loop
momenta corresponding to physical momenta  on the Fermi surface, and the second
term $G$ is once again regular as $q\to0$. The factor in the numerator arises 
because the kinematics restricts the integral over 3-momentum to $0\leq\vert{\bf l}\vert\leq\vert{\bf
q}\vert\cos\theta$.
 
The rest of the argument proceeds by examining the resulting integral equation
\begin{eqnarray}
\Gamma(p^\prime,p,q)=\Gamma^0(p^\prime,p)&-&
i\int{{d^4\ell}\over{(2\pi)^4}}\Gamma(p^\prime,\ell,q)G(\ell)\Gamma^0(\ell,p)
\nonumber\\
&+&{{{\mathfrak g}k_F^2}
\over{(2\pi)^3}}\int d\Omega\Gamma(p^\prime,\ell,q)\Gamma^0(\ell,p)
{{\vert{\bf q}\vert\cos\theta}\over{\omega-\vert{\bf q}\vert\cos\theta}}
\label{eq:gam}
\end{eqnarray}
in two distinct kinematic regimes as $q\to0$. First consider $\vert{\bf
q}\vert/\omega\to0$:
\begin{equation}
\Gamma^\omega(p^\prime,p)\equiv\lim_{q\to0,\,\vert{\bf q}\vert/\omega\to0}
\Gamma(p^\prime,p,q)=\Gamma^0(p^\prime,p)-i\int{{d^4\ell}\over{(2\pi)^4}}
\Gamma^\omega(p^\prime,\ell)G(\ell)\Gamma^0(\ell,p).
\label{eq:gamomeg1}
\end{equation}
In fact, it is possible to eliminate dependence on the regular functions
$\Gamma^0$ and $G$ by substituting (\ref{eq:gamomeg1}) in (\ref{eq:gam})
to yield
\begin{equation}
\Gamma(p^\prime,p,q)=\Gamma^\omega(p^\prime,p)+
{{{\mathfrak g}k_F^2}\over{(2\pi)^3}}\int
d\Omega\Gamma(p^\prime,\ell,q)\Gamma^\omega(\ell,p){{\vert{\bf
q}\vert\cos\theta}\over{\omega-\vert{\bf q}\vert\cos\theta}}.
\label{eq:gamomeg2}
\end{equation}
Now we can consider the opposite limit $\omega/\vert{\bf q}\vert\to0$ in
(\ref{eq:gamomeg2}):
\begin{equation}
\Gamma^k(p^\prime,p)\equiv\lim_{q\to0,\,\omega/\vert{\bf q}\vert\to0}
\Gamma(p^\prime,p,q)=\Gamma^\omega(p^\prime,p)-
{{{\mathfrak g}k_F^2}\over{(2\pi)^3}}\int 
d\Omega\Gamma^k(p^\prime,\ell)\Gamma^\omega(\ell,p).
\label{eq:gamk}
\end{equation}
The function $\Gamma^k$ describes forward scattering for quasiparticles located
on the Fermi surface, for which all physical 
scattering processes have $\omega=0$, 
and can be identified with $-{\cal A}$ calculated in the previous section.
In order to find a physical interpretation of $\Gamma^\omega$, we consider
(\ref{eq:gamomeg2}) in the vicinity of a pole of $\Gamma(p^\prime,p,q)$
considered as a function of $q$; in this case the first term on the RHS can be
neglected and the argument 
$p^\prime$ which plays a passive role can be suppressed. At the Fermi
surface the residual momenta $p$ and $\ell$ can be replaced by their
corresponding unit 3-vectors. With the definition  
\begin{equation}
\Phi(\hat{\bf p})={{\hat{\bf p}.{\bf q}}\over{\omega-\hat{\bf p}.{\bf q}}}
\Gamma(p,q)
\end{equation}
eqn. (\ref{eq:gamomeg2}) becomes an integral equation for excitations
propagating with wavevector {\bf q} and amplitude $\Phi$:
\begin{equation}
(\omega-\hat{\bf p}.{\bf q})\Phi(\hat{\bf p})=(\hat{\bf p}.{\bf q})
{{{\mathfrak g}k_F^2}
\over{(2\pi)^3}}\int d\Omega\Gamma^\omega(\hat{\bf l},\hat{\bf
p})\Phi(\hat{\bf l}).
\label{eq:zerosound}
\end{equation}
This equation\footnote{If proper account is taken of the absence of full
Lorentz invariance, then all 3-momenta in the argument 
leading to (\ref{eq:zerosound}) are rescaled by a factor $\beta_F$.}
describes collective excitations of the shape
of the Fermi surface known as zero sound, one of the principal
predictions of the
Fermi liquid theory \cite{Landau4}, with $\Gamma^\omega$ playing the role of
Fermi liquid interaction ${\cal F}$. Eqn. (\ref{eq:gamk}) thus yields the
full relation between ${\cal A}$ and ${\cal F}$: to lowest order in perturbation
theory we recover (\ref{eq:fl}).

\section{HDET Confronts the Fermi Liquid}
\label{sec:confront}
 
There are many situations in strong
interaction models, such as the chiral limit of the NJL model \cite{HKST}, or
the color symmetric channel in QCD \cite{BC}, where the direct contribution
to ${\cal A}$ and hence ${\cal F}$ vanishes. In this situation HDET would yield
a Fermi liquid interaction of the wrong sign, with potentially unphysical
consequences.

To see how this works, define the spin-symmetric Landau parameters
\begin{equation}
f_l^S\equiv(2l+1)\int{{d\Omega}\over{4\pi}}P_l(\cos\theta){\cal
F}_{{\bf k},{\bf k}^\prime}(\theta)
\end{equation}
where the integral ranges over all angles between ${\bf k}$ and ${\bf k}^\prime$
taken on the Fermi surface. I now quote without proof the following relations
for a relativistic Fermi liquid \cite{BC}:
\begin{eqnarray}
\beta_F&=&{k_F\over\mu}-{{{\mathfrak g}k_F^2}\over{6\pi^2}}{f_1^S};\\
{{\partial\mu}\over{\partial n}}&=& {{2\pi^2}\over{\mu{\mathfrak g}k_F}}
+f_0^S-{\textstyle{1\over3}}f_1^S.
\label{eq:compress}
\end{eqnarray}
Here $n={\mathfrak g}k_F^3/6\pi^2$ is the quark number density.
Generalising the exchange result (\ref{eq:qed})
to QCD with $N_c$ colors and $N_f$ flavors of massless quark (note the direct
term vanishes in the color symmetric channel) we have
\begin{equation}
f_0^S=g^2{{(N_c^2-1)}\over{8N_c^2N_f\mu^2}}\;\;;\;\;
f_1^S=0\;\;;\;\;{\mathfrak g}=2N_cN_f.
\end{equation}
Integrating (\ref{eq:compress}) we thus find to $O(g^2)$:
\begin{equation}
\beta_F={k_F\over\mu}=1-g^2{{(N_c^2-1)}\over{24N_c\pi^2}}.
\label{eq:flqed}
\end{equation}

For the HDET exchange interaction (\ref{eq:ahdet}), however, the corresponding
results are
\begin{equation}
f_0^S=f_1^S=-g^2{{(N_c^2-1)}\over{16N_c^2N_f\mu^2}}
\end{equation}
implying
\begin{equation}
\beta_F={k_F\over\mu}\left(1+g^2{{(N_c^2-1)}\over{48N_c\pi^2}}\right)
=1+g^2{{5(N_c^2-1)}\over{144N_c\pi^2}}.
\label{eq:flhdet}
\end{equation}
Thus the Fermi velocity is predicted to be superluminal in this approach.

At $O(g^2)$ we can correct for the mismatch between (\ref{eq:flqed}) and
(\ref{eq:flhdet}) by modifying the leading order HDET Lagrangian to
\begin{equation}
{\cal L}_{HDET}^\prime=\bar\psi_+(\gamma_0,c_1\hat{\bf p}\vec\gamma.\hat{\bf
p})^\mu i\tilde D_\mu\psi_+\;\;\;\mbox{with}\;\;\;
\tilde A_\mu=e^{-iX^\prime}A_\mu e^{iX^\prime}\;\;\;;\;\;\;X^\prime=c_2X,
\end{equation}
with the constants chosen to be 
\begin{equation}
c_1=1-g^2{{11(N_c^2-1)}\over{144N_c\pi^2}}\;\;\;;\;\;\;
c_2=1-g^2{{(N_c^2-1)}\over{18N_c\pi^2}}.
\end{equation}
However, whilst it is feasible to repair HDET to reproduce parameters directly
related to
the fundamental quanta, there are other phenomena where the failure to
deal with the exchange interaction correctly may be more serious. For instance,
in theories where the only non-trivial Landau parameter is $f_0^S$, 
eqn. (\ref{eq:zerosound}) may be solved for the zero sound velocity
$\beta_0=\omega/\vert{\bf q\vert}$ \cite{Landau4}:
\begin{equation}
\beta_0=\beta_F\left(1+2\exp\left(-{{4\pi^2\beta_F}
\over{{\mathfrak g}k_F^2f_0^S}}
\right)\right).
\end{equation}
The collective excitations are thus extremely sensitive to the
sign of $f_0^S$.

\section{Conclusion}

The main result of this paper is that if the $\psi_-$ degrees of freedom defined
by the projection (\ref{eq:decomp}) are eliminated, the resulting effective
theory incorrectly describes exchange interactions between quasiparticle 
states at the Fermi surface at tree level, 
as exemplified by the explicit perturbative 
QED calculation of
Sec.~\ref{sec:forward}. 

It should be stressed that this result does not invalidate the HDET programme;
the failure to reproduce exchange amplitudes
may be compensated by introducing a four-fermi contact term of $O(g^2/\mu^2)$, as
outlined by Sch\"afer \cite{Schafer}. In his approach all operators consistent
with the symmetries of the underlying gauge theory are included from
the start, and coefficients systematically
determined by matching at a suitably chosen scale 
$\Lambda\sim\mu$. 
Within the approach adopted here in which the leading order terms are
``derived'' via the projection (\ref{eq:unit}), the 
correction arises from integrating out $\psi_-$ states from the loop of
Fig.~\ref{fig:sdzero}, but with the naive power counting of $O(g^4/\mu^2)$
non-perturbatively enhanced to $O(g^2/\mu^2)$.

How serious a problem is this? It is worth
recalling that currently few people believe that quark matter is a Fermi liquid;
rather it is supposed to be a
 superconductor with energy gap $\Delta\sim O(10)$MeV at the Fermi surface.
The infra-red singularity leading to the
key relation (\ref{eq:gamk}) for the Fermi liquid interaction is thus
cut off, and
the dire consequences predicted in Sec.~\ref{sec:confront} no longer
hold. The
relativistic degenerate electrons in the interior of a white dwarf are
expected to form a Fermi liquid, but
in this case a photon screening mass
$\lambda\sim O(e\mu)$ should be included in the calculation, which implies
that the exchange amplitude (\ref{eq:qed}) is only $O(\alpha)$ compared to the
direct term. 

At the very least though, these considerations suggest that 
careful physical arguments
should be supplied before the HDET approach is applied, 
particularly since one of the main goals of any conceivable numerical
implementation would be to examine the parameter range of its applicability.

\section*{Acknowledgements}
The author is supported by a PPARC Senior Research Fellowship, and has greatly
benefitted from discussions with Deog Ki Hong, Steve Hsu, and Seyong Kim.

\end{document}